\documentclass[]{mn2e}  
\usepackage{mnras_cite}  
\newcommand{\nc}{\newcommand}    

\nc{\citealt}{\cite} 
\nc{\de}{\delta} 
\nc{\tISW}{\triangle_T^{ISW}}
\nc{\hn}{\hat{n}}
\nc{\bH}{\bar{H}} 
\nc{\Ol}{\Om_{\Lambda}} 

\nc{\ul}{\underline} \nc{\al}{\alpha} \nc{\g}{\gamma}
\nc{\Del}{\Delta} \nc{\e}{\textrm{e}} \nc{\eps}{\epsilon}
\nc{\lam}{\lambda} \nc{\Om}{\Omega} \nc{\Omm}{\Omega_m}
\nc{\Oml}{\Omega_\Lambda} \nc{\LCDM}{$\Lambda$CDM~} 
\nc{\ve}{\varepsilon} \nc{\mn}{{\mu\nu}} \nc{\vp}{\varphi}

\def\gsim{\; \raise0.3ex\hbox{$>$\kern-0.75em
\raise-1.1ex\hbox{$\sim$}}\; }

\nc{\be}[1]{\begin{equation}\mbox{$\label{#1}$}}    
\nc{\bea}[1]{\begin{eqnarray} \mbox{$\label{#1}$}}    
\nc{\Section}[2]{\section{#2}\label{#1}}    
\nc{\Bibitem}[1]{\bibitem{#1}}    
\nc{\Label}[1]{\label{#1}}    
    
\nc{\Mpc}{Mpc/h}    
\nc{\vev}[1]{\langle #1 \rangle}    
    
\nc{\eea}{\end{eqnarray}} \nc{\ee}{\end{equation}}  
\nc{\eeq}{\end{equation}}  
    
\def\lcdm{$\Lambda$CDM~}

\def\etal{{et al. }}    
\def\etals{{et al. }}    
\def\ltsima{$\; \buildrel < \over \sim \;$}    
\def\gtsima{$\; \buildrel > \over \sim \;$}    
\def\simlt{\lower.5ex\hbox{\ltsima}}    
\def\simgt{\lower.5ex\hbox{\gtsima}}    
\nc{\w}{$w_2(\theta)$\ }    
\nc{\ie}{i.e.}     
\nc{\eg}{e.g.}    
\def\q{{\hat n} }

\input epsf

\begin{document}

\title[Cross-correlation of WMAP 3rd year data and 
the SDSS DR4 galaxy survey]{Cross-correlation of WMAP 3rd year data  and 
the SDSS DR4 galaxy survey: new evidence for Dark Energy}

\author[Cabr\'e \etal]{A.Cabr\'e$^{1}$,  E.Gazta\~{n}aga$^{1,2}$,
M.Manera$^{1}$,  P.Fosalba$^{1}$ \& F.Castander$^{1}$\\  
$^{1}$Institut de Ci\`encies de l'Espai, CSIC/IEEC, Campus UAB,
F. de Ci\`encies, Torre C5 par-2,  Barcelona 08193, Spain\\
$^{2}$INAOE, Astrof\'{\i}sica, Tonantzintla, Puebla 7200, Mexico    
}

\maketitle  

\begin{abstract}     
We cross-correlate the third-year WMAP data with galaxy samples extracted from
the SDSS DR4 (SDSS4) covering 13\% of the sky, 
increasing by a factor of 3.7 the volume sampled
in previous analysis. The new measurements confirm a positive cross-correlation 
with  higher significance (total signal-to-noise of about 4.7). 
The correlation as a function of angular scale is well fitted by the
integrated Sachs-Wolfe (ISW) effect for LCDM flat FRW models 
with a cosmological constant. The combined analysis of different
samples gives $\Oml=0.80-0.85$ (68\% Confidence Level, CL)
or $0.77-0.86$ (95\% CL).  We find similar best fit values for $\Oml$  
for different galaxy samples with median redshifts of  
$z \simeq 0.3$ and $z \simeq 0.5$, indicating that the data scale
with redshift  as predicted by the LCDM  cosmology (with equation of 
state parameter $w=-1$). This agreement is not trivial, but can not yet 
be used to break the degeneracy constraints in the $w$ versus $\Oml$ 
plane using only the ISW data.

\end{abstract}    
    
    
\maketitle    
    
    
\section{Introduction}    
\label{sec:intro}    

Dark Energy (DE) models with late time cosmic acceleration, such as the     
$\Lambda$-dominated CDM model, predict a slow down for the  growth    
of 
the
linear gravitational potential at moderate redshift $z<1$, which    
can be observed as temperature anisotropies in the CMB: the so-called    
late integrated Sachs-Wolfe (ISW) effect. The ISW effect is expected to
produce an increase of power (a bump) in the amplitude of the  CMB fluctuations
at the largest scales, ie lower order multipoles, which are
dominated by cosmic variance. This expectation, seems challenged
by observations, as the first year WMAP results (WMAP1)  confirmed 
the low amplitude of the CMB quadrupole first measured by COBE (eg Hinshaw \etals 1996a). 
The discrepancy between the observations and the
\lcdm model is particularly evident in the temperature angular correlation 
function $w_2(\theta)$, 
which shows an almost complete lack of signal on angular scales $\theta \simgt 60$ degrees.
According to Spergel \etals (2003), the probability of finding such a
result in a spatially-flat \lcdm cosmology is about $1.5 \times
10^{-3}$. This was  questioned 
in Gazta\~naga \etal (2003) who found, using simulated \lcdm 
WMAP maps, a much lower
significance (less than 2-sigma) for $w_2(\theta)$. A low significance was 
also estimated by different studies 
(eg Efstathiou 2003, Olivera-Costa \etal 2003), although a 
discrepancy larger than 3-sigma still remains on both the 
quadrupole-octopole alignment 
(Tegmark, Oliera-Costa \& Hamilton 2003,  Olivera-Costa \etal 2003)
and the WMAP observed high value of the temperature-polarization
 cross-correlation on large scales (Dor\'e, Holder \& Loeb 2004).

\begin{figure*}    
{\centering    
{\epsfysize=6cm \epsfbox{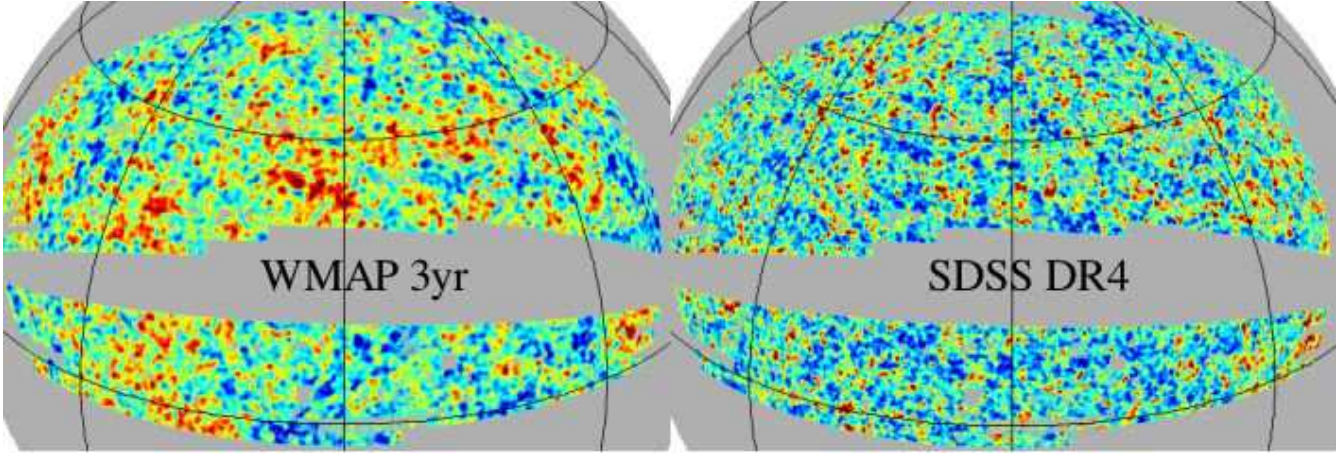}}    
}    
\caption{\label{fig:sdsswmap}    
SDSS DR4 galaxy density (LRG) fluctuation maps (right panel)  
compared to WMAP (V-band 3yr) temperature map (left panel). Both
maps are smoothed with a Gaussian beam of FWHM $=0.3$ deg.}
\end{figure*}    

Given the observed anomalies on the ISW predictions, 
it is of particular interest to check 
if the ISW effect can be detected observationally through an 
independent test, such as
the cross-correlation of temperature fluctuations with local tracers 
of the gravitational potential \cite{CT96}. 
A positive cross-correlation between WMAP1 and galaxy samples from the Sloan Digital Sky Survey 
(SDSS) was first found by Fosalba, Gazta\~naga \& Castander (2003, FGC03 from now on) 
and Scraton \etal (2003). FGC03
used the 1yr WMAP data (WMAP1) and the SDSS data release 1 (SDSS1). 
WMAP1 has also been correlated with the APM galaxies (Fosalba \& Gazta\~naga  2004),
infrared galaxies \cite{Afshordi}, 
radio galaxies \cite{Nolta}, and the
hard X-ray background \cite{BC04,BC04b}. 
The significance of these cross-correlations measurements was low (about 2-3 $\sigma$, 
see Gazta\~naga, Manera \& Multam\"aki 2006  for a summary and 
joint analysis),
and many scientists are still skeptical of the reality of these detections. 
Here we want to check if these results can be confirmed to higher significance using
the SDSS data release 4 (SDSS4) which covers $~3$ times the volume of SDSS1.
At the same time, we will compare the signal of the 1st and 3rd year WMAP data (WMAP3)
recently made public \cite{H06,Spergel06}.
With better signal-to-noise and better understanding of foreground contamination
in WMAP3, it remains to be seen whether 
the low significance signal of the WMAP1-SDSS1 analysis
can be confirmed with WMAP3-SDSS4, or if on the contrary this signal
vanishes as systematic and statistical errors are reduced.

\section{The Data}    
\label{sec:data}    
    
In order to trace the changing gravitational potentials we use
galaxies selected from the Sloan Digital Sky Survey Data Release 4
\cite{adelman06}, SDSS4 hereafter, which covers 6670
deg$^2$ (i.e, 16$\%$ of the sky). We have selected subsamples with
different redshift distributions to check the reliability of the
detection and to probe the evolution of the ISW effect. All subsamples
studied contain large number of galaxies, 
between 10$^6$-10$^7$, depending on
the subsample. We concentrate our analysis on the North Galactic Cap
SDSS4 Area ($\sim$ 5500 deg$^2$), because it contains the most
contiguous area. We have selected 3 magnitude subsamples with
$r=18-19$, $r=19-20$ and $r=20-21$ and a high redshift Luminous Red
Galaxy (LRG; e.g. Eisenstein \etal 2001) color selected subsample
($17<r<21$, $(r-i)>(g-r)/4 + 0.36$, $(g-r)> 0.72*(r-i)+1.7$). Because of
the smaller volume, the $r=18-19$ and $r=19-20$ subsamples provide low
signal-to-noise (S/N$<$2) in the cross-correlation with WMAP, and we
therefore center our analysis on the two deeper subsamples. 
The mask used for these data  avoids pixels with 
observed holes, trails, bleeding, bright 
stars or seeing greater than 1.8. 

To model the redshift distribution of our samples
we take a generic form of the type:
\be{eq:N(z)}
N(z) \sim \phi_G(z) \sim 
(z-z_c)^2 ~ \exp{{\left(-{{z-z_c}\over{z_0-z_c}}\right)^{3/2}}}, 
\ee
for $z>z_c$ and zero otherwise. 
The $N(z)$ distribution of the $r=20-21$ subsample
is quite broad with $z_c\simeq 0$ and $z_0 \simeq 0.2$ which
results in a median redshift,
$\bar{z} = 1.4 z_0 \simeq 0.3$ (e.g., Dodelson \etal 2001, Brown \etal 2003).
On the other hand, the LRG subsample has a  narrower redshift
distribution. The first colour cut is perpendicular to the 
galaxy
evolutionary tracks in the (g-r) .vs.(r-i) colour space and ensures
that very few $z<0.40$ galaxies are selected, which translates into 
a cut $z_c \simeq 0.37$ in the above $N(z)$ model.  The second colour cut
is parallel to evolution and perpendicular to spectral type
differences and selects only red galaxies with old stellar 
populations. The faint magnitude limit ($r<21$) cuts high redshift 
galaxies ($z_0 \simeq 0.45$), which results in an overall
median redshift of $\bar{z}\simeq0.5$.

We use the full-sky CMB maps from     
the third-year WMAP data \cite{H06,Spergel06} (WMAP3 from now on).
In particular, we have chosen the V-band ($\sim 61$ GHz)     
for our analysis since it     
has a lower pixel noise than the highest frequency W-band    
($\sim 94$ GHz), while it has sufficient high spatial resolution    
($21^{\prime}$) to map    
the typical Abell cluster radius at the mean SDSS depth. We use a combined
SDSS+WMAP mask that includes the Kp0 mask, which cuts $21.4 \%$    
of WMAP sky pixels \cite{B03b},    
to make sure Galactic emission does not affect our analysis.    
WMAP and SDSS data are digitized into $7^{\prime}$ pixels using the    
HEALPix tessellation     
\footnote{Some of the results in this paper have been     
derived using HEALPix \cite{GHW99},   
http://www.eso.org/science/healpix }.    
  Figure \ref{fig:sdsswmap} shows how the WMAP3 and SDSS4 pixel maps look like
when density and temperature fluctuations are smoothed on $0.3$ deg scale.

    
\section{Cross-Correlation and errors}    
\label{sec:cross}

\begin{figure}    
{\centering{
\epsfysize=7.cm \epsfbox{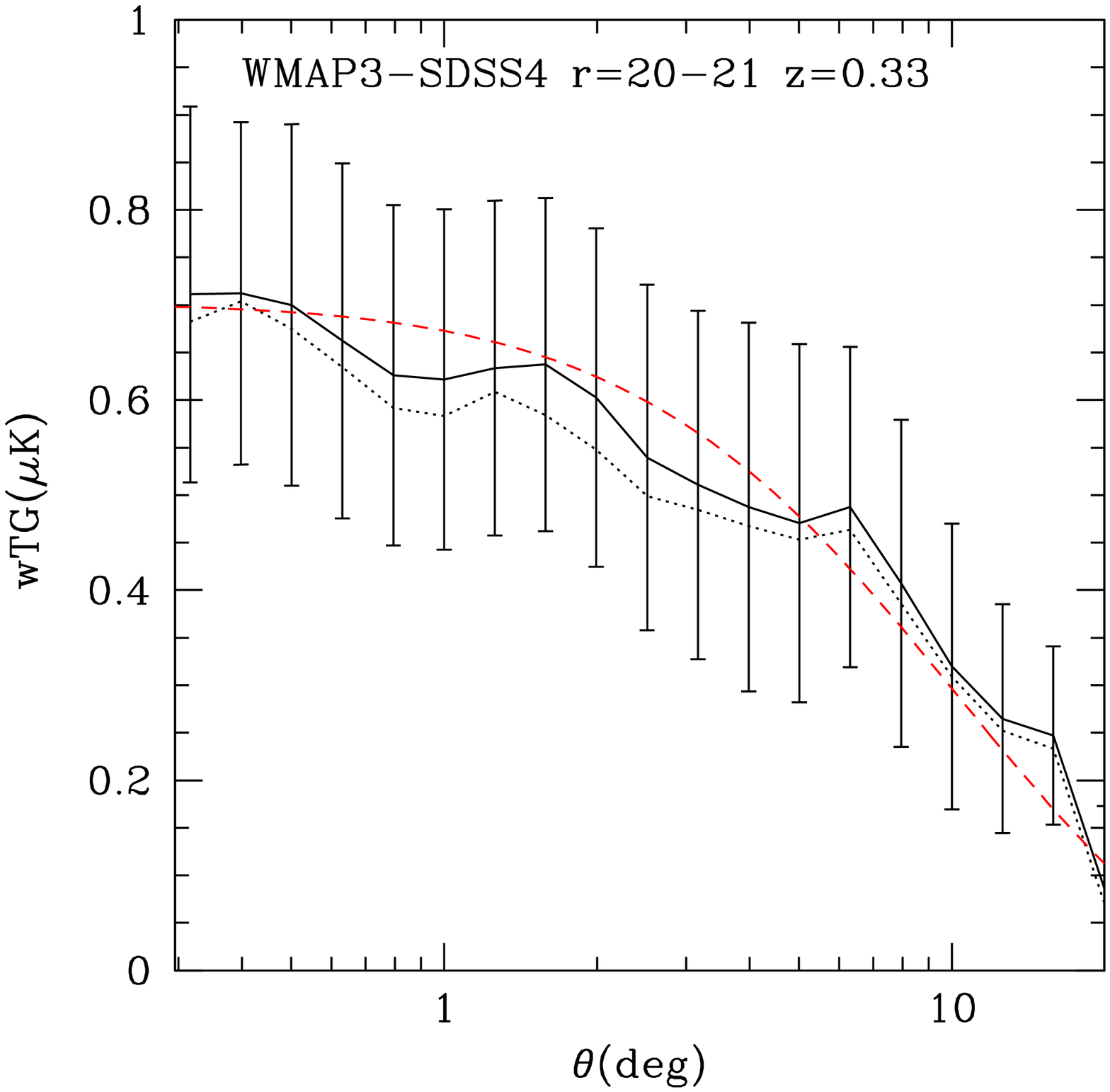} 
\epsfysize=7.cm \epsfbox{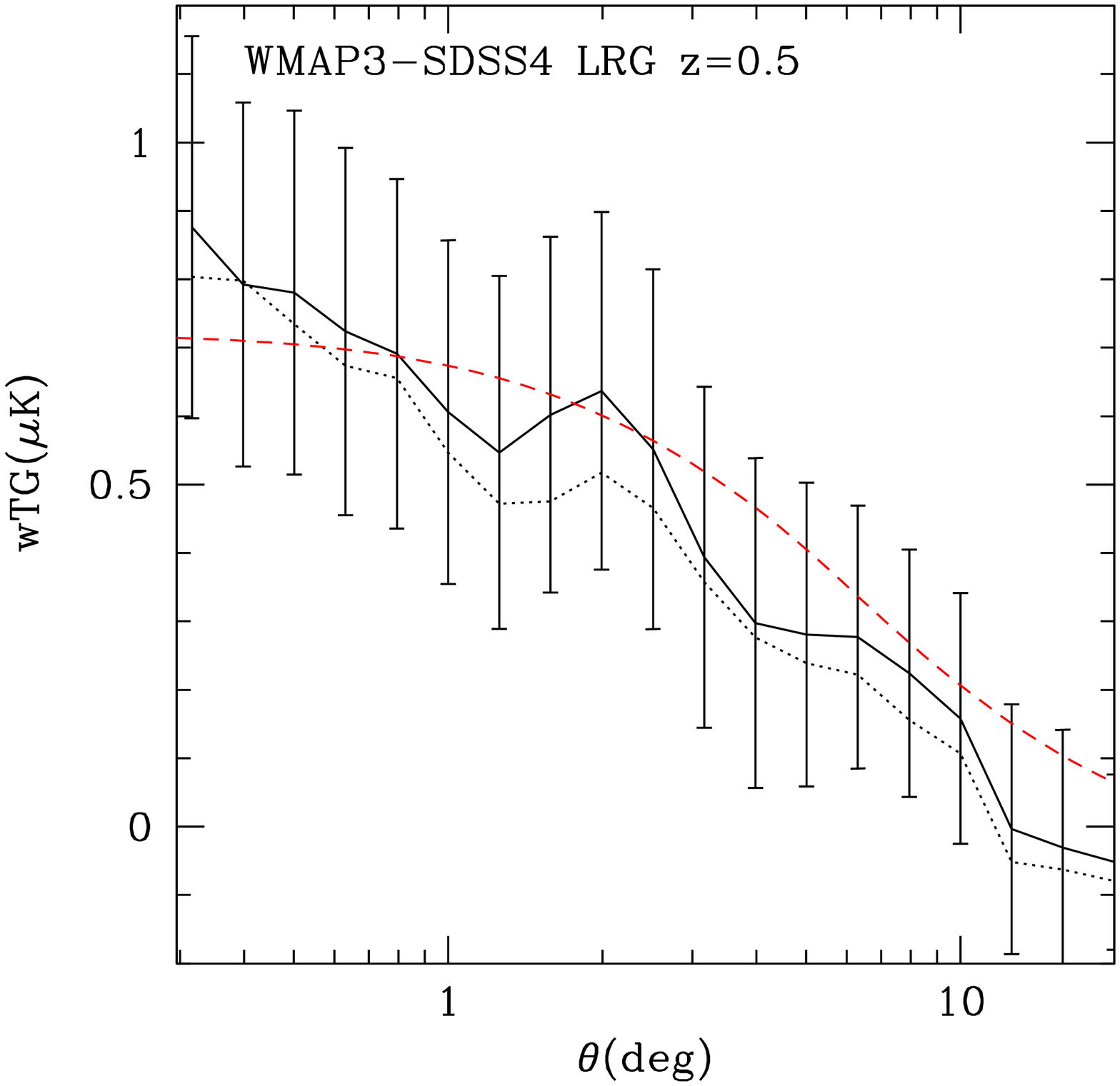}}}
\caption{\label{fig:wtg}     
The continuous line with errorbars shows the
 WMAP3-SDSS4 angular cross-correlation as a function of scale for the
 $r=20-21$ sample (top) and the LRG sample (bottom). 
The dotted line  corresponds to using 
the 1st yr WMAP (WMAP1-SDSS4) data, which is very close to the
WMAP3 results (continuous line).
The dashed lines show the \lcdm model with $\Oml =0.83$ ( best overall fit)
scaled to the appropriate bias and projected to each sample redshift.}
\end{figure}

We define the cross-correlation function as the expectation value of density    
fluctuations $\delta_G= N_G/<N_G>-1$ and temperature anisotropies    
$\Delta_T= T-T_0$ (in $\mu$K)    
at two positions $\q_1$ and $\q_2$ in the sky:    
$w_{TG}(\theta) \equiv  \vev{ \Delta_T({\bf\q_1}) \delta_G({\bf\q_2}) }$,    
where $\theta = |\bf{\q_2}-\bf{\q_1}|$, assuming that    
the distribution is statistically isotropic.    
To estimate $w_{TG}(\theta)$ from the pixel maps we use:    
\be{eqn:ctg}    
w_{TG}(\theta) = {\sum_{i,j} \Delta_T({\bf\q_i})~     
\delta_G({\bf\q_j}) ~w_i~w_j \over{\sum_{i,j} w_i~w_j}},    
\ee    
where the sum extends to all pairs $i,j$ separated by     
$\theta \pm \Delta\theta$.     
The weights $w_i$ can be used to minimize the variance when the pixel    
noise is not uniform, however this introduces larger cosmic variance.    
Here we follow the WMAP team and use uniform weights (i.e. $w_i=1$).     
The resulting correlation is displayed in Fig.\ref{fig:wtg}.
On scales up to $10$ degrees we find    
significant correlation above the estimated error-bars.  
The dotted and continuous lines correspond to WMAP1 and WMAP3 
data respectively, and show little difference within the errors. This
indicates that the cross-correlation is signal dominated. 
   
\begin{table}
\begin{center}
\begin{tabular}{|r|r|r|}
\hline
$\theta(deg)$&$wTG(20-21)$&$wTG(LRG)$\\
\hline
0.316 &   $0.711\pm0.198$ & $0.876\pm0.279$\\
0.398 &   $0.712\pm0.180$ & $0.793\pm0.266$\\
0.501 &   $0.700\pm0.190$ & $0.781\pm0.266$\\
0.631 &   $0.662\pm0.187$ & $0.724\pm0.268$\\
0.794 &   $0.626\pm0.179$ & $0.691\pm0.255$\\
1.000 &   $0.622\pm0.179$ & $0.606\pm0.251$\\
1.259 &   $0.634\pm0.176$ & $0.547\pm0.258$\\
1.585 &   $0.637\pm0.175$ & $0.602\pm0.260$\\
1.995 &   $0.603\pm0.178$ & $0.637\pm0.261$\\
2.512 &   $0.540\pm0.182$ & $0.552\pm0.263$\\
3.162 &   $0.511\pm0.183$ & $0.394\pm0.249$\\
3.981  &  $0.488\pm0.194$ & $0.298\pm0.241$\\
5.012  &  $0.470\pm0.188$ & $0.281\pm0.222$\\
6.310 &   $0.488\pm0.168$ & $0.277\pm0.192$\\
7.943  &  $0.407\pm0.172$ & $0.224\pm0.181$\\
10.000 &  $0.320\pm0.150$ & $0.158\pm0.183$\\
12.589  & $0.265\pm0.121$ & $-0.004\pm0.182$\\
15.849 &  $0.247\pm0.094$ & $-0.031\pm0.172$\\
19.953 &  $0.086\pm0.087$ & $-0.052\pm0.128$\\
\hline
\end{tabular}
\caption{$w_{TG}(\theta)$ for WMAP3-SDSS4.}
\end{center}
\label{taula}
\end{table}

We have used different prescriptions to estimate the
covariance matrix:
a) jack-knife, b) 2000 montecarlo simulations
c) theoretical estimation
(including cross-correlation signal) both in configuration and harmonic
space. Our montecarlo simulations in b) include independent simulations 
of both the CMB and galaxy maps, with the adequate cross-correlation signal. 
All three estimates give very similar results for covariance and the $\chi^2$ errors,
details will be presented elsewhere (Fosalba \etal 2006). 

To compare models we use a $\chi^2$ test:
\be{eq:chi}    
\chi^2 = \sum_{i,j=1}^{N} \Delta_i ~ C_{ij}^{-1} ~ \Delta_j,    
\ee    
where $\Delta_i \equiv w_{TG}^E(\theta_i) - w_{TG}^M(\theta_i)$    
is the difference between the "estimation" $E$ and the    
model $M$. We perform a Singular Value Decomposition (SVD) of the covariance
matrix $C_{ij}=(U_{ik})^\dagger D_{kl}V_{lj}$ where $D_{ij}= \lambda_i^2 \delta_{ij}$ 
is a diagonal matrix with the singular values on the diagonal, and $U$ and $V$ 
are orthogonal matrices that span the range and nullspace of $C_{ij}$.
We can choose the number of eigenvectors $\widehat{w}_{TG}(i)$ (or principal components)
 we wish to include in our $\chi^2$ by 
effectively setting the corresponding inverses of the small singular values to zero. 
In practice, we work only with the subspace of ``dominant modes" which have a significant
``signal-to-noise'' (S/N). The S/N of each eigenmode, labeled by $i$, is:
    
\be{StoN}
\Big(\frac{S}{N}\Big)_i =  \Big| \frac{\widehat{w}_{TG}(i)}{\lambda_i} \Big|= 
 \Big| \frac{1}{\lambda_i}\sum_{j=1}^{N_b} U_{ji} \, \frac{w_{TG}(j)}{\sigma_w(j)} \Big|.
\ee
As S/N depends strongly on the assumed cosmological  model, we use the
direct measurements of $w_{TG}$ to estimate this quantity.
The total  S/N can be obtained by adding the individual modes in quadrature. 
In our analysis we have used 5 eigenmodes for the $r=20-21$ sample
and 3 for the LRG sample. The results are similar if we use less eigenmodes. 
With more eigenmodes, the inversion becomes unstable because we include eigenvalues
which are very close to zero and are dominated by noise.

\subsection{Comparison with Predictions}    
\label{sec:isw}    

ISW temperature anisotropies are given by \cite{SW67}: 
\be{ISW1} 
\tISW(\hn) \equiv {T(\hn)-T_0\over{T_0}} =- 2 \int dz ~{d\Phi\over{dz}}(\hn,z)
\ee 
where $\Phi$ is the Newtonian gravitational potential at redshift
$z$.  One way to detect the ISW effect is to cross-correlate
temperature fluctuations with galaxy density fluctuations projected in
the sky \cite{CT96}. 
It is useful to expand the cross-correlation 
$w^{ISW}_{TG}(\theta)  = { <\tISW(\hn_1) \de_G(\hn_2)>}$
in a Legendre polynomial basis. On large linear scales and
small angular separations it is:

\bea{final_wtg}
w^{ISW}_{TG}(\theta) & =  &
\sum_l {2l+1\over 4\pi}\, p_{l}(\cos\theta)\, C_{GT}^{ISW}(l) \nonumber \\
C_{GT}^{ISW}(l) & = & {4 \over (2l+1)^2}
\int dz\, W_{ISW}(z) W_G(z) {H(z)\over c} P(k) \nonumber\\
W_{ISW}(z) & = & 3 \Omm (H_0/c)^2  {d[D(z)/a]\over dz} \\
W_G(z) & = &  b(z)\phi_G(z) D(z),\nonumber
\eea
where $k={l+1/2\over r}$, $\phi_G(z)$ is the survey galaxy selection 
function in Eq.[\ref{eq:N(z)}] and  $r(z)$ is the comoving distance. 
This is just a Legendre decomposition of the equations presented
in Fosalba \& Gazta\~naga (2004), see also Afshordi (2004).
The advantage of this formulation 
is that we can here set the monopole ($l=0$)
and dipole ($l=1$) contribution to zero, as it is  done in the WMAP maps. The contribution
of the monopole and dipole  to $w_{TG}$ is significant and  over predicts  $w_{TG}$ by
about 10\%. The power spectrum is
$P(k)= A~k^{n_s}~T^2(k)$, where 
$T(k)$ is the \LCDM transfer function, which we evaluate using
the fitting formula 
of Einseintein \& Hu 1998.

We make the assumption that on very large scales the galaxy distribution is
a tracer of the underlaying  matter fluctuations, related through the linear
bias factor, $\de_G(\hn,z)=b(z)\de_m(\hn,z)$. We estimate $b(z)$
from the angular  galaxy-galaxy auto-correlation $w_{GG}(\theta)$ in each sample by
fitting to the linear flat \lcdm model prediction  $w_{GG}(\theta)$ 
 and marginalizing over the value of $\Omega_m$. The models
have $h=0.71$, $T_{CMB}=2.725$, $\Omega_B=0.022/h^2$, $n_s=0.938$ and $\Omega_k=0$.
and are normalized
to the value of $\sigma_8$ that best fits WMAP3 data (Spergel \etal 2006):  
$\sigma_8 = 0.75 \pm^{0.03}_{0.04}$. 
With this procedure we find a normalization of 
$b \sigma_8 \simeq 0.90-0.96$ and  $b \sigma_8 \simeq 1.02-1.12$  for the
 $r=20-21$ and LRG samples respectively. We marginalize all our results over
 the uncertainties in both $\sigma_8$ and $b\sigma_8$. This also roughly accounts for the
uncertainty in the selection function. The predictions of $w_{TG}$ do no change much
with the selection function (see \S4.1 in  Gazta\~naga, Manera \& Multam\"aki 2006),
but the bias estimated from
from $w_{GG}$ depends strongly on the effective volume covered by $\phi_G(z)$. Because
of the marginalization our final results do not change much when we change the median
redshift of the sample by $\sim 10\%$, which represents current uncertainties
in $N(z)$. But in the case of the LRG it is critical to include not only the correct
value of the mean redshift (or $z_0$ in Eq.[\ref{eq:N(z)}]) but also the redshift
cut $z_c$ introduced by the color selection  in Eq.[\ref{eq:N(z)}].
In previous LRG cross-correlation analysis 
(eg FGC03 \& Gazta\~naga, Manera \& Multam\"aki 2006)
the value of $z_c$ was neglected. This can over predict
$b \sigma_8$, as estimated from $w_{GG}$, by a factor of two. 
Uncertainties in the shape of $N(z)$ considered here are within 
the normalization errors we have already included for $\sigma_8$ and $b\sigma_8$.
We have also made predictions for the best fit WMAP3 data with $n_s=1$
which gives different parameters and normalization ($\sigma_8 = 0.79 \pm^{0.05}_{0.06}$)
and find very similar results.

\begin{figure}    
\centering{    
\epsfysize=7.cm \epsfbox{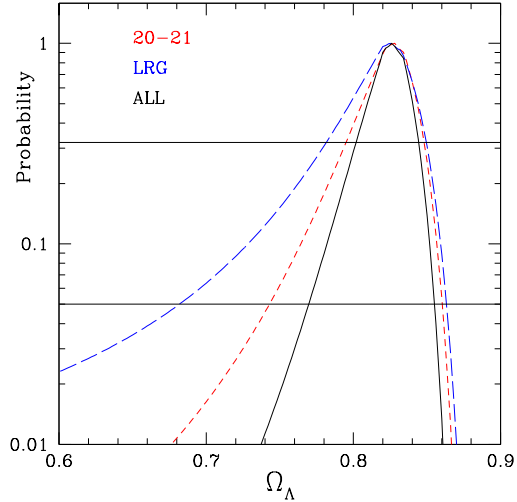}}
\caption{\label{fig:pdf}    
Probability distribution: $1-P_\chi[>\Delta\chi^2,\nu=1]$
for $\Oml$ in the $r=20-21$ sample (short-dashed
line), the LRG sample (long-dashed line) and the combined analysis 
(continuous middle curve). The range of $68\%$ and $95\%$ confidence regions
in $\Oml$ are defined by the intersection with the corresponding horizontal lines.}
\end{figure}

Under the above assumptions we are left with only one free parameter, 
which is $\Omega_m$ or $\Omega_\Lambda = 1-\Omega_m$. 
 Fig.\ref{fig:pdf} shows the probability distribution
estimated for $\Omega_\Lambda$ from the $\Delta \chi^2=\chi^2-\chi^2_{min}$ analysis 
away from the minimum value $\chi^2_{min}$. Both samples 
prefer the same value of $\Oml$. This is a consistency check for the
$\Lambda CDM $ model.
The combined best fit model has $\Oml \simeq 0.83 ^{+0.02}_{-0.03}$.
The predictions for this $\Oml$ best value are shown as a dashed line in 
Fig. \ref{fig:wtg}.

\begin{figure}    
\centering{    
\epsfysize=6.cm \epsfbox{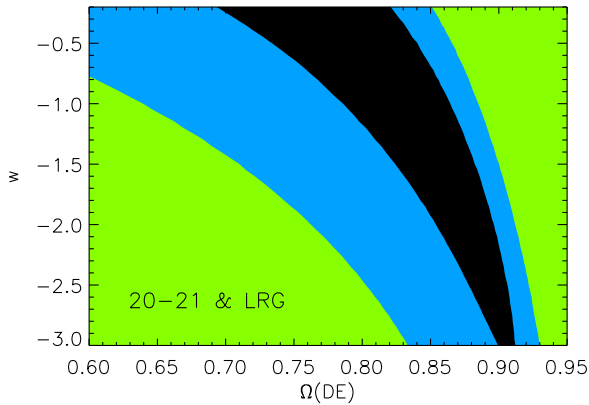}}
\caption{\label{fig:wx}    
Two dimensional contours
 for  $\Oml$ and $w$, the DE effective equation of state.
The inner black contour limits the 1D marginalized $68\%$
confidence region ($\Delta\chi^2=1$). The other contour 
correspond to $95\%$ limits ($\Delta\chi^2=4$).}
\end{figure}

Fig.\ref{fig:wx} shows the joint  2D contours for dark energy models
with an effective equation of state $w=p/\rho$, assuming not
perturbations in DE and a Hubble
equation: $H^2/H_0^2 = \Om (1+z)^3+ \Oml (1+z)^{3(1+w)}$. 
For each $(w,\Oml)$ we derive $b\sigma_8$ consistently from the 
galaxy-galaxy auto-correlation data. We also marginalize over
the uncertainties in $b\sigma_8$ and over $\sigma_8 \in 
 (0.65, 0.85)$, to account for the WMAP3 $\sigma_8$ normalization
for $w \neq -1$.
The cosmological constant model $w=-1$, however, still remains 
a very good  fit to the data. This is due to the large degeneracy of the 
equation of  state  parameter $w$ with $\Oml$. This degeneracy can be broken by
supernovae SNIa data (eg see  Corasaniti, Giannantonio and Melchiorri 2005
and Fig.8 in see Gazta\~naga, Manera \& Multam\"aki 2006).

\section{Discussion}    
\label{sec:discuss}

The objective of our analysis was primarily to check if we could confirm or
refute with higher significance the findings of the WMAP1-SDSS1 cross-correlation
by FGC03. With an increase in area of a factor of $\simeq 3.7$ in SDSS4, larger
 signal-to-noise and better understanding of foregrounds in WMAP3, 
our new analysis shows that the signal is robust. 
This is also  in line with the first
findings using optical (APM) galaxies  by Fosalba \& Gazta\~naga (2004).
The cross-correlation signal in  WMAP3-SDSS4 seems slightly larger 
than in previous WMAP1-SDSS1 measurements 
which results in slightly larger values for $\Oml$ (see FGC03 and
Gazta\~naga, Manera \& Multam\"aki 2006).
This is probably due to sampling variance, as the SDSS4 volume
has increase by almost a factor of 4 over SDSS1. We find little difference within
the errors in the cross-correlation of WMAP1-SDSS4 and WMAP3-SDSS4 (see Fig.\ref{fig:wtg}).

The total S/N in Eq.\ref{StoN}  of the WMAP3-SDSS4 correlation is
$S/N \simeq 3.6$ for the $r=20-21$ sample and $S/N \simeq
3.0$ for the LRG, which gives a combined $S/N \simeq 4.7$,
assuming the two samples are independent. 
We have checked  the validity of this  assumption by 
doing a proper join analysis where we include the covariance between
the two samples. For the join analysis we find a $S/N \simeq 4.4$ with the
first 2 dominant eigenvectors and $S/N \simeq 4.8$ with 4 eigenvectors.

We find that a $\Lambda CDM$ model with $\Ol\simeq 0.83$ successfully 
explains the ISW effect for both samples of galaxies without need
of any further modeling.  The best fit for $\Ol$ for each 
individual sample are very close.
This is significant and can be understood as a consistency test
for the \lcdm model.

The equation of state parameter appears to be very degenerate and 
it is not well constrained
by current ISW data alone (see also Corasaniti, Giannantonio and Melchiorri 2005  and
Gazta\~naga, Manera \& Multam\"aki 2006). 
Upcoming surveys such as the Dark Energy Survey (DES, www.darkenergysurvey.org), 
with deeper galaxy samples and more accurate redshift information should be able 
to break the $w-\Oml$ degeneracy and maybe shed new light on the  
the nature of dark energy.

\section*{Acknowledgments}  
    
We acknowledge the support from Spanish Ministerio de Ciencia y    
Tecnologia (MEC), project AYA2005-09413-C02-01  with EC-FEDER funding
and research project 2005SGR00728 from  Generalitat de Catalunya.
AC and MM aknowledge support from the DURSI department of the Generalitat 
de Catalunya and the European Social Fund. PF acknowledges support 
from the Spanish MEC through a Ramon y Cajal fellowship.
This work was supported by the European Commission's ALFA-II programme
through its funding of the Latin-american European Network for
Astrophysics and Cosmology (LENAC).

\end{document}